hep-ph/0411319
\documentclass[12pt]{iopart}
\usepackage{iopams}
\usepackage{graphicx}
\begin{document}
%%%
\title{Quasi-particle model of strongly interacting matter}
%%%
\author{Marcus Bluhm\dag\,, Burkhard K\"ampfer\dag\, and Gerhard Soff\ddag}
%%%
\address{\dag\ Forschungszentrum Rossendorf, PF 510119, 01314 Dresden,
Germany}
\address{\ddag\ Institut f\"ur Theoretische Physik, TU Dresden, 01062
  Dresden, Germany}
%%%
\ead{\mailto{M.Bluhm@fz-rossendorf.de}}
%%%
\begin{abstract}
  The successful quasi-particle model is compared with recent lattice
  data of the coefficients in the Taylor series expansion of
  the excess pressure at finite temperature and baryon density. A chain of
  approximations, starting from QCD to arrive at the model expressions
  for the entropy density, is presented.
\end{abstract}
%%%
%\pacs{..., ..., ...}
%%%
%\submitto{\jpg}
%%%
%\maketitle
%%%%%%%%%%%%%%%%%%%%%%%%%%%%%%%%%%%%%%%%%%%%%%%%%%%%%%%%%%%%%%%%%%%%%%%%%%%%%%
\section{Introduction}

The equation of state of strongly interacting matter at finite
temperature $T$ and small chemical potential $\mu$ has become accessible
fairly recently through first principle lattice QCD
calculations~\cite{All03,Csi04,Fod02,Fod03}. Apart from solving QCD on the
lattice, there exist analytical approaches such as resummed HTL
scheme, $\Phi$ functional approach etc. (cf.~\cite{Ris04} for a
recent survey). Ab initio approaches~\cite{Bla99,Bla02} are 
restricted in describing lattice data on temperatures 
$T\gtrsim 2.5T_{\rm c}$, where $T_{\rm c}$ is the transition
temperature of deconfinement and chiral symmetry restoration. In contrast,
phenomenological models with adjustable
parameters~\cite{Pes94,Pes00,Lev98} cover the region $T\gtrsim T_{\rm
  c}$. Here, we present new developments of our quasi-particle model
(QPM)~\cite{Pes94,Pes00}. 

In~\sref{sec:model} the QPM is reviewed. In~\sref{sec:coeffs} the
model is confronted with recent lattice QCD data. Motivating our
model,~\sref{sec:QCD}
briefly illustrates a chain of approximations within a $\Phi$ funtional
approach. The results are summarized in~\sref{sec:conclusions}. 

\section{Quasi-particle model}
\label{sec:model}

The model is based on the idea that the quark-gluon fluid can be
expressed in terms of quasi-particles. The pressure of $N_q$
light (q), strange (s) quarks and gluons (g) reads  
\begin{equation}
  \label{equ:pres}
  \fl
  p(T,\mu)=\sum_{i\,=\,q,s,g} p_i(T,\mu) - B(T,\mu) .
\end{equation}
$p_i$ are thermodynamic standard expressions in which $T$
and $\mu$ dependent self-energies $\Pi_i$ enter. $B(T,\mu)$, together
with the stationarity condition $\delta p/\delta m_i^2=0$~\cite{Gor95},
ensures thermodynamic self-consistency (cf.~\cite{Pes00,Pes02} for
details). Thus, the entropy density $s=\partial p/\partial
T=\sum_{i\,=\,q,s,g} s_i$ explicitly reads
\begin{equation}
  \label{equ:entr}
  \fl
  s_i=\frac{d_i}{2\pi^2T}\int_0^\infty \rmd
  k\,k^2\left\{\frac{\left(\frac{4}{3}k^2 +
  m_i^2\right)}{\omega_i(k)}\left[f_i^+(k) + f_i^-(k)\right] -
  \mu\left[f_i^+(k)-f_i^-(k)\right]\right\}
\end{equation}
with statistical distribution functions $f_i^\pm (k)=(\exp
[(\omega_i(k)\mp\mu_i)/T]+S_i)^{-1}$, $S_q=S_s=1$, $S_g=-1$,
$\mu_q=\mu$, $\mu_s=\mu_g=0$ and degeneracies $d_q=6N_q$, $d_s=6$ and
$d_g=8$. 

In the thermodynamically relevant region of momenta $k\sim
T$, $\mu$, the quasi-particle dispersion relations are approximated by
the asymptotic mass shell expressions near the light cone
$\omega^2_i(k)=k^2 + m_i^2$ with $m_i^2=\Pi_i(k;T,\mu) +
(x_iT)^2$. The self-energies $\Pi_i$ of the quasi-particle excitations are
approximated by their 1-loop expressions at hard
momenta~\cite{Pes00} neglecting imaginary parts, and $x_iT$
represent the quark masses used on the 
lattice~\cite{All03}. Replacing the running coupling in the
self-energies by an effective coupling, $G^2(T,\mu)$, non-perturbative
effects are thought to be taken into
account~\cite{Pes02}. Imposing thermodynamic consistency onto $p$, a flow equation
for $G^2(T,\mu)$ follows~\cite{Pes00} 
\begin{equation}
  \label{equ:flow}
  \fl
  a_\mu\frac{\partial G^2}{\partial\mu} + a_T\frac{\partial
  G^2}{\partial T} = b
\end{equation}
which can be solved as Cauchy problem by knowing $G^2$ on an arbitrary curve
$T(\mu)$. One convenient choice is parametrizing $G^2(T(\mu=0))$
appropriately (cf.~\cite{Blu04}) such that $p$ and $s$ can be computed at
non-vanishing $\mu$ from~(\ref{equ:pres},\ref{equ:entr}). 

\section{Expansion coefficients}
\label{sec:coeffs}

Apart from~(\ref{equ:pres}), the pressure can be decomposed into a
Taylor series 
\begin{equation}
  \label{equ:series}
  \fl
  \frac{p(T,\mu)}{T^4}=\frac{p(T,\mu=0)}{T^4} + \frac{\Delta
  p(T,\mu)}{T^4} = \sum_{n=0}^\infty
  c_{2n}(T)\left(\frac{\mu}{T}\right)^{2n} 
\end{equation}
with $c_0(T)=p(T,\mu=0)/T^4$ and vanishing $c_k$ for odd $k$. The
expansion coefficients $c_k$ have been subject of recent lattice
evaluations~\cite{All03} by computing derivatives of the thermodynamic
potential. They follow from~(\ref{equ:pres}) as
$c_k(T)=\left.\partial^k p/\partial\mu^k\right|_{\mu=0}T^{k-4}/k!$ in
our model and read 
\begin{eqnarray}
  \label{equ:coeff2}
  \fl
    c_2 & = \frac{3N_q}{\pi^2T^3}\int_0^\infty \rmd k\, k^2
    \frac{\rme^{\omega_q/T}}{\left(\rme^{\omega_q/T}+1\right)^2} \, , \\
  \label{equ:coeff4}
  \fl
    c_4 & = \frac{N_q}{4\pi^2T^3}\int_0^\infty \rmd k\, k^2
    \frac{\rme^{\omega_q/T}}{\left(\rme^{\omega_q/T}+1\right)^4}
    \left\{\rme^{2\omega_q/T} - 4\rme^{\omega_q/T} +1
    -\frac{T}{\omega_q}\left[\rme^{2\omega_q/T}-1\right]A_2\right\}\,,
\end{eqnarray}
where $A_2=(G^2/\pi^2+3x_q\sqrt{G^2/(6\pi^4)}+
[\frac{3}{2}x_qT^2/\sqrt{6G^2}+\frac{1}{2}T^2]\partial^2 
G^2/\partial\mu^2)|_{\mu=0}$, $\omega_q$ is taken at
$\mu=0$ and $\partial^2G^2/\partial\mu^2$ follows from
differentiating~(\ref{equ:flow}). 

\begin{figure}[h]
\begin{center}
\includegraphics[width=.5\textwidth]{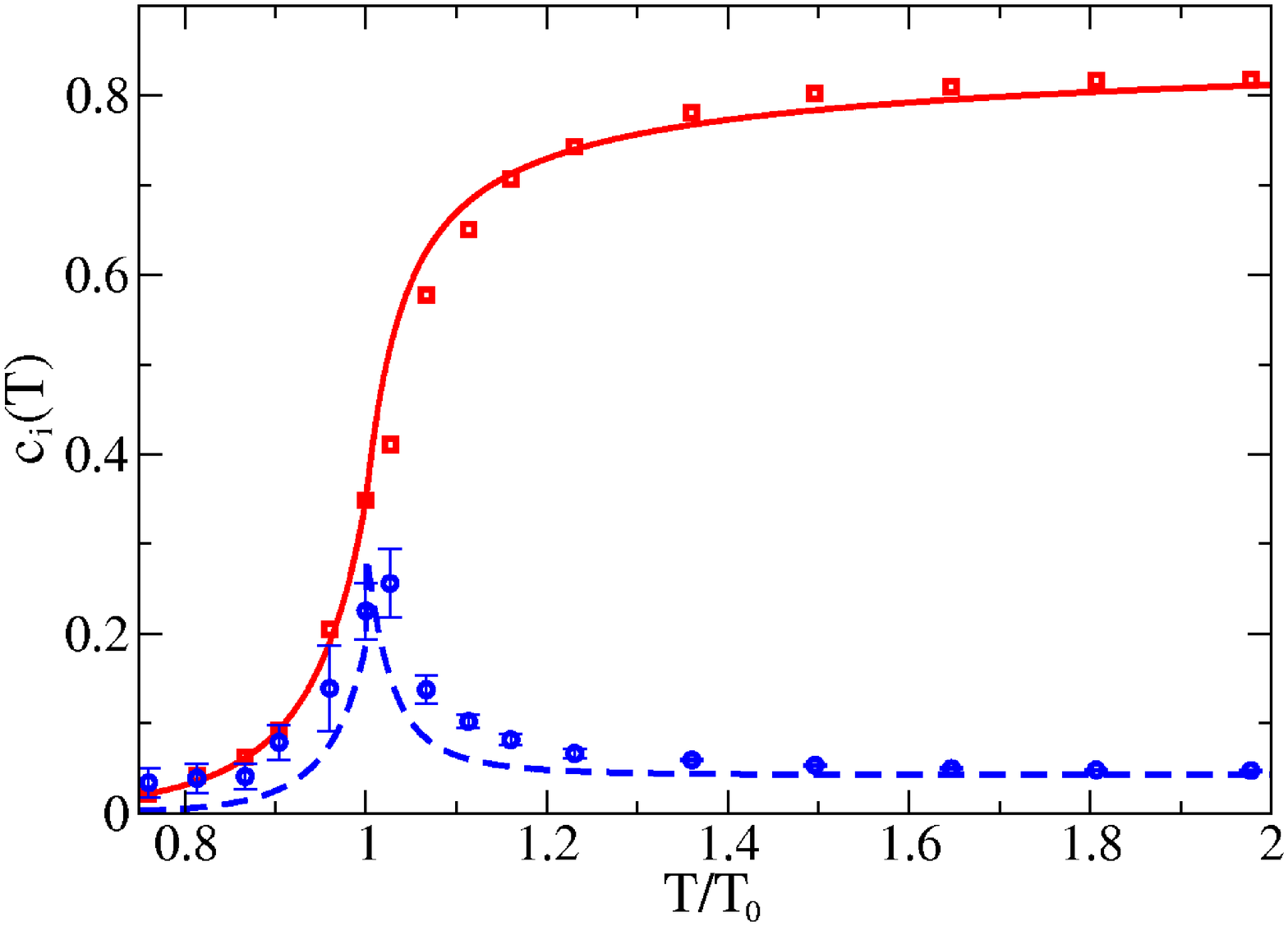}
\includegraphics[width=.48\textwidth]{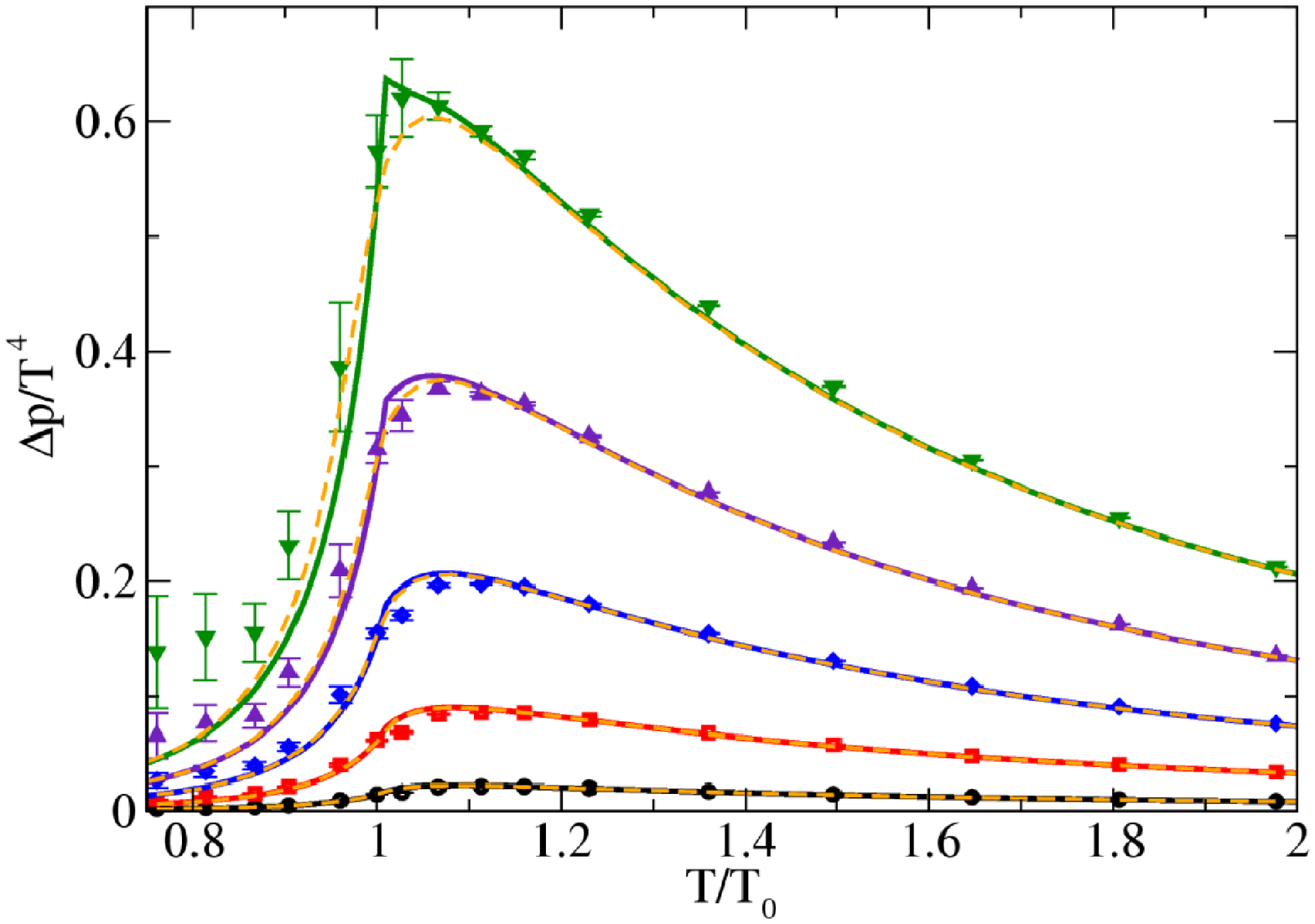}
%\parbox[h]{.7\textwidth}{
  \caption{\label{fig:coeffsexcess} Left: Expansion coefficients $c_2(T)$
    (squares) and $c_4(T)$ (circles). Data from~\cite{All03}. Full
    and dashed curves represent, respectively, corresponding QPM
    results~(\ref{equ:coeff2},\ref{equ:coeff4}). Right: Scaled excess
    pressure $\Delta p$ from~\cite{All03} (symbols)
    and QPM results (full lines) from truncating~(\ref{equ:series}) at
    order $(\mu/T)^4$ for different 
    $\mu/T_0=0.2,0.4,0.6,0.8,1.0$ (lower to upper curve). Dashed
    lines depict corresponding full QPM results.}
%}
\end{center}
\end{figure}

In the left panel of~\Fref{fig:coeffsexcess}, the QPM results of $c_{2,4}$ 
calculated from~(\ref{equ:coeff2},\ref{equ:coeff4}) are compared with
lattice QCD results~\cite{All03} for the two-flavour case,
i.e. $N_q=2$. Using $x_q=0.4$ and $x_g=0$ 
as in~\cite{All03} and setting $T_{\rm c}(\mu=0)\equiv T_0=170$ MeV,
$G^2(T(\mu=0))$ is adjusted to describe $c_2(T)$. Since
$c_4$ in~(\ref{equ:coeff4}) only depends on $G^2$ and its derivatives
at $\mu=0$, no further assumptions enter into the evaluation once the
parametrization is fixed. A fairly good agreement is found for
$c_4(T)$. Note, in particular, that the pronounced peak structure of
$c_4$ at $T_0$ solely originates from the term including $\left.\partial^2
G^2/\partial\mu^2\right|_{\mu=0}$. 

In~\cite{All03}, the excess pressure $\Delta p$ has been
calculated as the truncation of the expansion~(\ref{equ:series})
including the order $(\mu/T)^4$. The right panel
of~\Fref{fig:coeffsexcess} exhibits the 
comparison of $\Delta p$ calculated by employing only $c_{2,4}$ 
in~(\ref{equ:series}) (full lines) with the lattice data for different 
$\mu/T_0$. An impressively good agreement with the data 
is observed for small values of $\mu$ in which case $c_4$ is of less
importance. Similarly, $\Delta p$ can be evaluated as infinite series
from the QPM by combining~(\ref{equ:pres}) and~(\ref{equ:series})
(dashed lines). These full results
differ noticeably from the truncated results only for $T\approx
T_0$. It should be noted that the model is successfully applied to 
describing the equation of state with strange quarks~\cite{Sza03}. 

\section{Contact with QCD}
\label{sec:QCD}

Motivating the strong assumptions made in formulating the QPM
expressions~(\ref{equ:pres},\ref{equ:entr}), a chain of reasonable
approximations starting from QCD would be of desire. The thermodynamic
potential $\Omega=-pV$ in ghost free gauge reads~\cite{Bla99} 
\begin{equation}
\fl
  \Omega[D,S] = T\left\{\frac{1}{2}{\Tr}\left[\ln D^{-1}-\Pi
  D\right] - {\Tr}\left[\ln S^{-1}-\Sigma S\right]\right\} +
  T\Phi[D,S] 
\end{equation}
with dressed propagators $D$ and $S$ of bosons and fermions and
corresponding exact self-energies $\Pi$ and $\Sigma$ from Dyson's
equations. The functional $\Phi$ is given by the sum over all 2
particle irreducible skeleton diagrams and $\Pi$, $\Sigma$ follow from
truncating dressed propagator lines in these
diagrams~\cite{Bla99}. The sum over Matsubara frequencies in the trace
$\Tr$ is transformed into an appropriate contour integral in the
complex energy plane. Computing the entropy density $s=-\partial
(\Omega/V)/\partial T$, an ultra-violet finite expression 
\begin{eqnarray}
  \label{equ:phientropy}
  \nonumber
  \fl
  s = &-{\rm tr}\int\frac{d^4k}{(2\pi)^4}\frac{\partial
  n(\omega)}{\partial T} [{\rm Im}\ln D^{-1}-{\rm Im}\,\Pi\,{\rm Re}
  D] \\
  \fl
  &- 2{\rm\, tr}\int\frac{d^4k}{(2\pi)^4}\frac{\partial
  f(\omega)}{\partial T} [{\rm Im}\ln S^{-1}-{\rm Im}\,\Sigma\,{\rm
  Re} S] + s'
\end{eqnarray}
is derived. After truncating $\Phi$ at 2 loop order, $s'=0$ is 
found. Lost gauge invariance gets restored by employing hard thermal
loop expressions for $\Pi$ and $\Sigma$ which show the correct limiting
behaviour for $k\sim T$, $\mu$. Performing the remaining trace tr over discrete
indices in~(\ref{equ:phientropy}), quantum numbers of quarks and gluons are
recovered. Furthermore, the exponentially suppressed longitudinal
gluon modes and the plasmino branch can be neglected. In addition,
neglecting imaginary parts in $\Pi$ and $\Sigma$ as well as Landau
damping and approximating self-energies and dispersion relations
suitably, expression~(\ref{equ:entr}) for the entropy density $s$ is
recovered. 

\section{Conclusion}
\label{sec:conclusions}

The quasi-particle model has been reviewed and successfully compared
with recent lattice data of the expansion coefficients $c_{2,4}$ and
the excess pressure at finite temperature and chemical 
potential. Briefly, a chain of approximations within the $\Phi$
functional scheme starting from QCD has been summarized which leads to
our model. 

%%%
\ack
Inspiring discussions with A~Peshier are gratefully acknowledged. The
work is supported by BMBF, GSI and EU-I3HP. 
%%%

%%%
\end{document}